\documentclass[twocolumn, longbibliography,floatfix]{revtex4-1}

\usepackage{soul}
\usepackage{amsmath}   
\usepackage{amssymb}
\usepackage{mathtools}
\usepackage{graphicx}
\usepackage{dcolumn}
\usepackage{bm}
\usepackage{amsmath}
\usepackage{graphicx}
\usepackage{amsfonts}
\usepackage{subfigure}
\usepackage{graphicx}
\usepackage{array}
\usepackage{float}
\usepackage{color}
\usepackage{amssymb}%
\usepackage[colorlinks=true,linkcolor=blue]{hyperref}%
\hypersetup{allcolors=blue}
\usepackage[normalem]{ulem}
\usepackage{xcolor}

\newcommand{\bra}[1]{\ensuremath{\left\langle #1 \right\vert}}
\newcommand{\ket}[1]{\ensuremath{\left\vert #1 \right\rangle}}
\usepackage{mathtools}
\DeclarePairedDelimiterX\braket[2]{\langle}{\rangle}{#1 \delimsize\vert #2}

\begin{document}

\title{Gauge effects in bound-bound Rydberg-transition matrix elements}
\date{\today }

\author{A. Duspayev}
    \email{alisherd@umich.edu}
\affiliation{Department of Physics, University of Michigan, Ann Arbor, MI 48109, USA}
\author{G. Raithel}
\affiliation{Department of Physics, University of Michigan, Ann Arbor, MI 48109, USA}

\begin{abstract}
Accurate data on electric-dipole transition matrix elements (EDTMs) for bound-bound Rydberg-atom transitions become increasingly important in science and technology. Here we compute radial EDTMs of rubidium using the length, velocity and acceleration gauges for electric-dipole-allowed transitions between states with principal and angular-momentum quantum numbers ranging from 15 to 100. Wave-functions are computed based upon model potentials from Marinescu et al., Phys. Rev. A {\bf{49}}, 982 (1994). Length-gauge EDTMs, often used for low-$\ell$ transitions, are found to deviate from the fundamentally more accurate velocity-gauge EDTMs by relative amounts of up to $\sim 10^{-3}$. We discuss the physical reasons for the observed gauge differences, explain the conditions for applicability of the velocity and length gauges for different transition series, and present a decision tree of how to choose EDTMs. Implications for contemporary Rydberg-atom applications are discussed.
\end{abstract}

\maketitle
\section{Introduction}
\label{sec:intro}

Rydberg atoms~\cite{gall} are an active field of modern physics with applications in precision measurements~\cite{haroche, tan, ramos, berl}, molecular physics~\cite{shafferreview, feyreview}, quantum control~\cite{Moore2019a, cardman2020}, field sensing~\cite{Sedlacek2012, Holloway2014, RFMS, anderson2021}, nonlinear quantum optics~\cite{chang, liang, cantu}, and as a platform for quantum computing and simulations~\cite{Saffman.2016, browaeysreview, morgado2021}. Being at the core of several directions in fundamental research and emerging quantum technologies~\cite{Adams_2019}, Rydberg transitions from low-lying states or other Rydberg atomic states require accurate calculations~\cite{safronova2004, vsibalic2017arc}. Of particular interest are electric and magnetic multipole transition matrix elements~\cite{safronova2011}, static and dynamic polaizabilities~\cite{safronova2004, Chen2015}, and collisional and photo-ionization~\cite{dinneen1992, markert2010, Viray2021, Cardman2021} cross-sections. Among the former, electric-dipole transition matrix elements (EDTMs) are the most important due to their wide usage~\cite{safronova2004}. 

A common framework for the computation of EDTMs between different bound Rydberg levels is based on model potentials~\cite{Marinescu}. In most cases, the computations are performed in the length gauge (LG)~\cite{bethe, friedrichbook}, also referred to as Babushkin gauge~\cite{papoulia}. Another common form is the velocity gauge (VG). In addition, there exists an acceleration gauge (AG). Although quantum mechanics is gauge-invariant~\cite{bethe, sakurai}, different approximations and assumptions must be made to transform the expressions for EDTMs between different gauges, which can result in notable discrepancies in the final results~\cite{Grant1974, Froese2009}. These are revealed in high-precision calculations and naturally raise the question which gauge should be used. This issue has been discussed in a range of research fields, including interaction of atoms and molecules with strong fields~\cite{leone, Cormier, bauer, chen2009, han2010, Bandrauk}, solid-state physics~\cite{virk, passos} and astrophysical spectroscopy~\cite{refId0, papoulia}.

\begin{figure}[htb]
 \centering
 \includegraphics[width=0.48\textwidth]{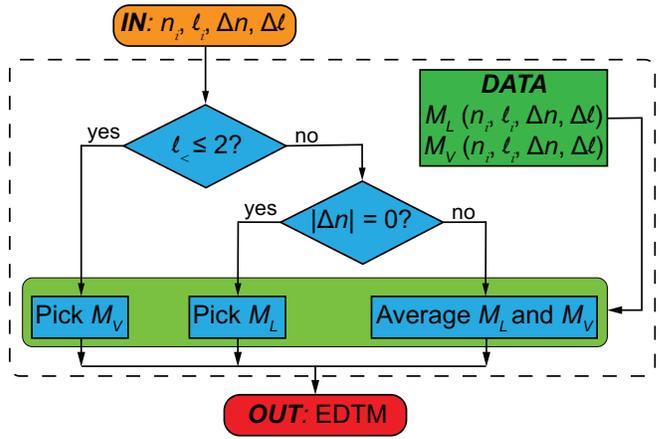}
  \caption{Decision tree to select EDTMs from pre-computed length-gauge and velocity-gauge databases of EDTMs, $M_L (n_i, \ell_i, \Delta n, \Delta \ell)$ and $M_V (n_i, \ell_i, \Delta n, \Delta \ell)$, respectively. Initial- and final-state principal and angular-momentum quantum numbers are denoted $(n_i, \ell_i)$ and $(n_f, \ell_f)$, respectively, the difference $\Delta n = n_f - n_i$, the difference $\Delta \ell = \ell_f - \ell_i = \pm1$, and $\ell_<$ is the lesser of $\ell_i$ and $\ell_f$.}
  \label{fig1}
\end{figure}

To our knowledge, gauge effects in EDTMs have not been broadly discussed yet in context with emerging applications of Rydberg-atom transitions in the aforementioned quantum technologies. Here we provide a comparison of EDTMs computed in the LG, VG and AG gauges for electric-dipole electromagnetic transitions in rubidium Rydberg atoms. The main emphasis is placed on the LG and VG forms, the most widely used and the most accurate gauges, respectively, while the AG is discussed for instructive purposes. We show that the angular-momentum type of the Rydberg-Rydberg transition plays a decisive role in what is the best choice for the gauge that should be used. For the Rb model potentials~\cite{Marinescu} that we use to compute the Rydberg-electron wave-functions, we find that the VG should be used if the lesser of the involved angular momenta, $\ell_<$, is 2 or lower, and that the VG or the LG can be used otherwise, as they produce identical results within the numerical accuracy level of the calculations. As an exception, only the LG should be used for transitions between near-degenerate Rydberg levels with $l_< \ge 3$. The decision tree that follows is summarized in Fig.~\ref{fig1} and is rationalized throughout the paper. While the AG should not be used in applications, for instructive purposes we show that the AG results converge with the VG and LG results in the appropriate limits.

The paper is organized as follows: The three gauge forms are reviewed and the relevant expressions are provided in Sec.~\ref{sec:gauges}, the results are presented in Sec.~\ref{sec:results} and discussed in Sec.~\ref{sec:disc}. The paper in concluded in Sec.~\ref{sec:concl}.

\maketitle
\section{Matrix elements in different gauge forms}
\label{sec:gauges}
In this section, we review the formalism for calculations of EDTMs in the three gauges, and justify the validity of the electric-dipole approximation (EDA). We begin with a spin-less, non-relativistic $N$-electron atom with nuclear charge $Z$ placed in a plane-wave electromagnetic field. The linearly polarized field has a vector potential $\mathbf{A} ({\bf{r}}, t) = {\bf{A}}_0 \cos({\bf{k}} \cdot {\bf{r}} - \omega t)$, wavevector $\mathbf{k}$, angular frequency $\omega$, and an amplitude of the electric field of $E_0 = \omega A_0$. The atom can be described by the following Hamiltonian:

\begin{eqnarray}
\hat{H} = \hat{H}_0 + \hat{H}_{int},
\label{Htotal}
\end{eqnarray}

\noindent where the field-free part, $\hat{H}_0$, is (in SI units) 

\begin{eqnarray}
\hat{H}_0 = \sum_{i=1}^N \left(\frac{\hat{\mathbf{p}}_i^2}{2m_e} - \frac{Z e^2}{4 \pi \epsilon_0 \hat{r}_i}\right) + \frac{1}{4 \pi \epsilon_0} \sum_{\substack{i=1\\i\neq j}}^N \frac{e^2}{|\hat{\mathbf{r}}_i - \hat{\mathbf{r}}_j|},
\label{H0}
\end{eqnarray}

\noindent and the atom-field interaction part, $\hat{H}_{int}$, resulting from including the kinetic momentum, $\hat{\mathbf{P}}_i = \hat{\mathbf{p}}_i + e \hat{\mathbf{A}} (\hat{\mathbf{r}}_i, t)$, is

\begin{equation}
\begin{split}
\hat{H}_{int} = \sum_{i=1}^N \Bigl( \frac{e}{2m_e} [ \hat{\mathbf{p}}_i \cdot \mathbf{A} (\hat{\mathbf{r}}_i, t) + \mathbf{A} (\hat{\mathbf{r}}_i, t) \cdot \hat{\mathbf{p}}_i] \\
+ \frac{e^2 \mathbf{A}^2 (\hat{\mathbf{r}}_i, t)}{2 m_e} \Bigr). 
\end{split}
\label{Hint}
\end{equation}

\noindent Here, $e$ is the magnitude of the electron charge, $m_e$ is the electron mass, and $\hat{\mathbf{r}}_i$ are the position operators of the electrons. The ${\bf{A}}^2$-term in Eq.~\ref{Hint} describes the ponderomotive interaction, which can be employed to realize ponderomotive optical lattices for Rydberg atoms~\cite{dutta2000, Anderson2011}. In the present work, we focus on the first term, which describes, among other phenomena, Rydberg-atom microwave transitions.

We consider an alkali atom with a single electron excited to Rydberg state. The sum in Eq.~\ref{Hint} can be dropped, and the interactions of the Rydberg electron with the core electrons can be compounded into a set of $\ell$-specific model potentials~\cite{Marinescu}. Adopting the Coulomb gauge, $\nabla \cdot \mathbf{A} = 0$, in which $\hat{\mathbf{p}}$ and $\mathbf{A} (\hat{\mathbf{r}}_i)$ commute, and also dropping the $\mathbf{A}^2$-term, which is irrelevant for the low-field transitions considered here, Eq.~\ref{Hint} can be written as:

\begin{equation}
\hat{H}_{int} = \frac{e}{m_e} \mathbf{A} (\hat{\mathbf{r}}) \cdot \hat{\mathbf{p}}. 
\label{Hintn}
\end{equation}

\noindent We then express the electric-field-normalized atom-field interaction matrix element between initial, $\ket{i}$, and final, $\ket{f}$, states of the Rydberg atom, in the rotating-wave approximation, as:

\begin{align*}
M_V = \frac{\bra{f} \hat{H}_{int} \ket{i}}{(E_0/2)} = - \frac{e \hbar}{{\rm{i}} m_e \omega} (\hat{\mathbf{n}} \cdot \int \psi_f^* e^{{\rm{i}} \mathbf{k}\cdot\mathbf{r}} \nabla \psi_i d^3 r) \quad.
\end{align*}

\noindent Making the electric-dipole approximation (EDA), which is valid for microwave transitions of Rydberg atoms and is addressed further below, 

\begin{equation}
M_V = - \frac{e \hbar}{{\rm{i}} m_e \omega} (\hat{\mathbf{n}} \cdot \int \psi_f^*~\nabla \psi_i~d^3 r),
\label{MV}
\end{equation}

\noindent where $\hat{\mathbf{n}}$ is the electric field's polarization vector. The electric-dipole transition matrix element (EDTM) in Eq.~\ref{MV} is commonly referred to as the velocity(VG)-form, as it essentially involves the linear-momentum operator. In order to transform it into the often-used length-gauge(LG) form, one invokes the commutation relation $[\hat{\mathbf{r}}, \hat{H}_0] = \frac{i \hbar}{m_e} \hat{\mathbf{p}}$, which is exact only if $\hat{H}_0$ does not contain momentum-dependent potentials~\cite{bethe}. For the case of an on-resonant interaction, $\omega = \vert E_f - E_i \vert / \hbar$ with final- and initial-state energies $E_f$ and $E_i$, respectively, the EDTM takes the commonly employed LG form,

\begin{equation}
M_L = e (\hat{\mathbf{n}} \cdot \int \psi_f^*~\mathbf{r}~\psi_i~d^3 r).
\label{ML}
\end{equation}

For our third, the acceleration-gauge (AG) form, to be valid, the valence-electron potential in $\hat{H}_0$ must be of pure Coulomb form, $\hat{V} = - Z e^2 / (4 \pi \epsilon_0 \hat{r})$, such that the commutation relation $[\hat{\mathbf{p}}, \hat{H}_0] = -i \hbar \nabla \hat{V}$ is true~\cite{bethe}. Then, the EDTM can be expressed in AG form,

\begin{equation}
M_A = -\frac{Z e^3}{m_e \omega^2} (\hat{\mathbf{n}} \cdot \int \psi_f^*~\frac{\hat{\mathbf{r}}}{4 \pi \epsilon_0 r^3}~\psi_i~d^3 r).
\label{MA}
\end{equation}

\noindent Note that the EDTMs in Eqs.~\ref{MV}-\ref{MA} are in SI units. In order to convert them to atomic units used in the next Section, one needs to divide by $e a_0$, where $a_0$ is the Bohr radius. (Alternately, set $\hbar = m_e = Z = e = 4 \pi \epsilon_0 = 1$ and convert $\omega$ from SI into atomic units.)

The VG is the most general form, whereas for the LG and AG to be applicable specific requirements must be satisfied. These always hold in the non-relativistic hydrogen atom. However, the alkali model potentials~\cite{Marinescu}, which we use here, are $\ell$-dependent for $\ell \leq 3$, which makes the LG inaccurate at these low $\ell$-values. Rydberg atoms with a polarizable ionic core, such as Rb and Cs, exhibit a long-range, non-Coulombic core polarization potential, the leading term of which scales as $- \alpha_d/(2 r^4)$. There, $\alpha_d$ is the dipolar core polarizability. Due to the non-Coulombic perturbation, the AG is inaccurate at most $\ell$-values.  It is an objective of our work to quantify the deviations of EDTMs computed in the LG and the marginally applicable AG forms from those computed in the fundamentally accurate VG form. 

It is noteworthy that the EDA must be applied to the VG, the fundamental gauge that follows directly from the ${\bf{A}} \cdot {\bf{p}}$-part of the minimal-coupling Hamiltonian in Eq.~\ref{Hintn}, before the matrix element can be transformed into the LG and AG forms (under the applicable respective conditions). For the microwave transitions considered in the present work, the EDA is naturally satisfied due to the long radiation wavelength (${\bf{k}} \cdot {\bf{r}} \ll 1$). It has been shown elsewhere that the EDA even applies to electric Rydberg-atom couplings in the UV wavelength range~\cite{Cardman2021}, where the Rydberg-atom size typically exceeds the wavelength  (${\bf{k}} \cdot {\bf{r}} > 1$).

In the numerical results presented in 
Sec.~\ref{sec:results}, we show comparisons between the EDTMs in the three gauges for several selected angular-momentum channels. 
All EDTMs shown are radial parts; matrix elements for transitions with specific magnetic quantum numbers follow from multiplication with angular matrix elements provided, for instance, in~\cite{bethe}.

The fine structure of the Rydberg levels is ignored in our analysis, as it would primarily only add Clebsch-Gordon factors to the angular matrix elements. The gauge effects we focus on in our work result from the $\ell$-dependence and the non-Coulombic long-range terms in the model potentials~\cite{Marinescu} used to compute the radial wave-functions of the $(n, \ell)$-states of the atom. 
To capture gauge effects, it is therefore sufficient to consider radial EDTMs with fine-structure-averaged quantum defects~\cite{gall} for the Rydberg energy levels. As the model potentials vary if $\ell\in[0,3]$ and are identical if $\ell \ge 3$, for $\ell_< \le 2$ we expect to find errors of both the LG and AG results relative to the VG result. The latter is considered correct within our model. For $\ell_< \ge 3$ we still expect to find errors of the AG result due to the non-Coulombic, long-range ion-core polarization potential.

For the numerical calculation of the required wave-functions, we use a method described in~\cite{Reinhard2007} that allows a non-uniform spatial grid. The spatial finesse is essentially characterized by a specifiable number of grid points per local de-Broglie wavelength of the electron wave-function (which varies widely within the atomic potential). For the bulk of our calculations, we chose a lower limit of 5,000 grid points per electron wavelength. To estimate numerical uncertainty, several computation series were performed on finer grids with a minimum of 10,000 grid points per electron wavelength. All computations were performed in Fortran using REAL*16 precision. For reference, less than 500 grid points per electron wavelength, in a REAL*8 implementation, suffice to compute EDTMs of strong bound-bound Rydberg transitions with three to four significant digits.
  
\maketitle
\section{Results}
\label{sec:results}
We have computed the EDTMs in all three gauges for all electric-dipole-allowed Rydberg-Rydberg transitions $n_i \rightarrow n_f$, with initial, $n_i$, and final, $n_f$, principal quantum numbers covering the full range from 15 to 100, and for all combinations of initial and final angular momenta, $\ell_i$ and $\ell_f$, that satisfy the electric-dipole selection rule $\Delta \ell = \ell_f - \ell_i = \pm 1$. The results are stored in data banks $M_V (n_i, \ell_i, \Delta n, \Delta \ell)$, $M_L (n_i, \ell_i, \Delta n, \Delta \ell)$, and $M_A (n_i, \ell_i, \Delta n, \Delta \ell)$, with $\Delta n$ defined as $\Delta n  = n_f - n_i$.

As a convenient measure for the differences between the radial EDTMs in the three gauges, we define the relative gauge error

\begin{align*}
\delta_{i,V} = \left|1 - \frac{M_i}{M_V}\right|, i = L, A, 
\end{align*}

\noindent where the subscripts V, L and A stand for VG, LG and AG gauges, respectively.

\begin{figure}[t!]
 \centering
  \includegraphics[width=0.48\textwidth]{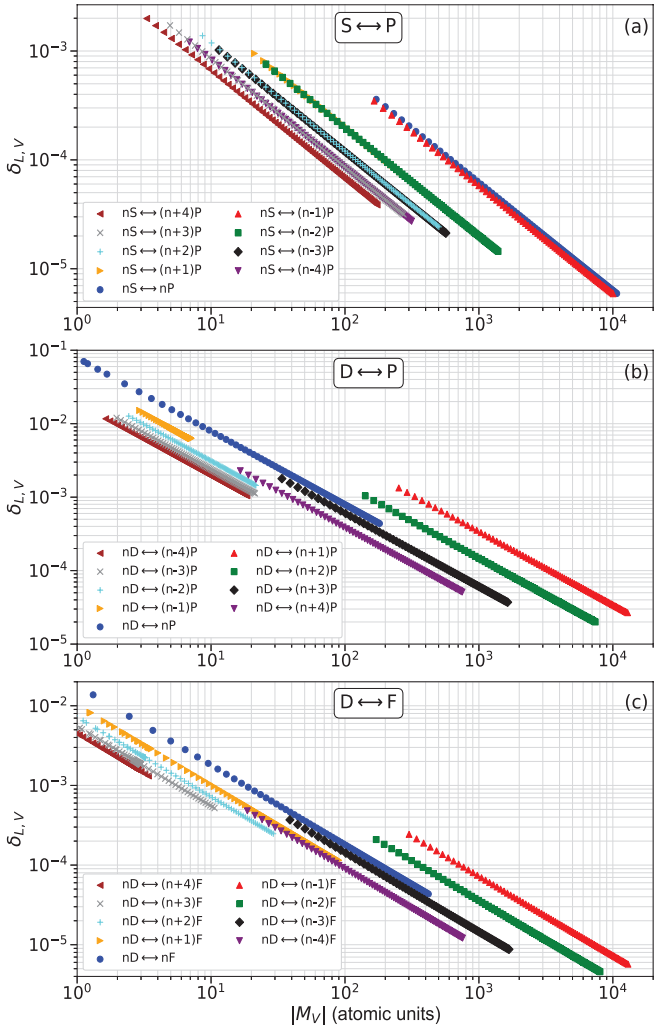}
  \caption{(Color online) Comparison of EDTMs calculated in LG vs VG using Eqs.~\ref{MV} and~\ref{ML} for the indicated $n_i S \leftrightarrow n_f P$ series (a), $n_i D \leftrightarrow n_f P$ series (b), and $n_i D \leftrightarrow n_f F$ series (c).The principal quantum numbers $n$ range between 15 (left margin, in most series), and 100 (right margin, in most series).}
  \label{fig2}
\end{figure}

In Fig.~\ref{fig2}, we show $\delta_{L, V}$ vs $|M_V|$ on double-log scales for several transition series with $\ell_< \le 2$. We consider transitions with $|\Delta n| \leq 4$, which include the series with the largest EDTMs (that are relevant in numerous applications). It is first noted that each series falls on a well-defined line characteristic to the series. All characteristic lines have similar slopes, with the distributions of data points $(|M_V|, \delta_{L, V})$ along those lines showing some irregularity for weak transitions (transitions with small $|M_V|$).  

Rb $n_i S_{1/2}$ $\leftrightarrow$ $n_f P_{J}$ microwave transitions are of some importance because the $n_i S_{1/2}$ Rydberg atoms can be initialized by two-photon laser excitation from the Rb $5S_{1/2}$ ground state. As seen in Fig.~\ref{fig2}(a), for the strongest transitions, $nS \leftrightarrow nP$ and $nS \leftrightarrow (n-1)P$, $\delta_{L,V}$ typically ranges between $10^{-4}$ and $10^{-5}$, while for the weaker transitions with larger $|\Delta n|$ this number can be as large as $\sim 2\times 10^{-3}$. Further, from the observed approximate scaling $\delta_{L,V} \propto 1 / |M_V|$ it follows that $\Delta_{L,V} : = M_L - M_V$ tends to be a constant for each series. For the strongest series, $nS \leftrightarrow nP$ and $nS \leftrightarrow (n-1)P$,  it is $\Delta_{L,V}  \approx -0.062$~ea$_0$ and $-0.058$~ea$_0$, respectively. For the weaker series, the approximately constant values of $\Delta_{L,V}$ generally drop. For instance, for the $nS \leftrightarrow (n+4)P$ series it is $\Delta_{L,V}  \approx -0.0066$~ea$_0$.  

Rb $n_i D$ $\leftrightarrow$ $n_f P$ microwave transitions are often used because of the large values of $|M_V|$ on the $nD$ to $(n+1)P$ series, 
and because the two-photon laser preparation rates of $n_i D$ Rydberg atoms from the Rb $5S_{1/2}$ ground state are greater than those of $n_i S$ Rydberg atoms.
It is seen in Fig.~\ref{fig2}(b) that these transitions have $\delta_{L, V}$-values on the order of three times larger than those of comparable  $n_i S$ $\leftrightarrow$ $n_f P$ transitions, with numerous instances of $\delta_{L,V} > 10^{-2}$. 
On the $n_i D \leftrightarrow n_f P$ series, the $\delta_{L, V}$ are much larger and the $M_V$ much smaller for $\Delta n \le 0$ than for $\Delta n > 0$.
The strongest series, $nD \leftrightarrow (n+1)P$, has $M_L - M_V \approx -0.34$~ea$_0$. 
Interestingly, the $nD \leftrightarrow nP$ series is very weak and exhibits a peculiar behavior at small $|M_V|$. We attribute the unusual features of the $nD \leftrightarrow nP$ series to the exact level energies and wave-functions of the involved states, which depend on the Rb quantum defects of both initial and final levels. In fact, further check revealed that this series exhibits what may be characterized as a bound-bound analog of a Cooper minimum~\cite{cooper1962, Viray2021}. (A Cooper minimum~\cite{cooper1962} is a class of EDTM zeros at certain quantum-defect-dependent photo-ionization energies.) In the present case, the $nD \leftrightarrow nP$ EDTMs pass through a pronounced minimum of $|M_V| \sim 0.16$~ea$_0$ at $n=20$ (off-scale in Fig.~\ref{fig2}(b)), which is about three orders of magnitude lower than the $|M_V|$-value of the nearby $20D \leftrightarrow 21P$ transition (which has $|M_V| = 472$~ea$_0$). In Fig.~\ref{fig2}(b), this results in an irregularity of the distribution of data points on the $nD \leftrightarrow nP$ series at low $|M_V|$. Since the relation $\delta_{L,V} \propto 1 / |M_V|$ still holds, $\delta_{L,V}$ reaches as high as $\sim 0.5$ for the $20D \leftrightarrow 20P$ transition.  Further examination reveals another unusual series, the $nD\leftrightarrow (n-1)P$ series. We found that in that case the EDTMs pass through a shallow maximum of $|M_V|$ of only $\approx 7$~ea$_0$ at $n \sim 65$. This maximum results in a ``bunching" of data points in Fig.~\ref{fig2}(b) at the right margin of the $nD \leftrightarrow (n-1)P$ series.

We further show several $n_i D \leftrightarrow n_f F$ series in Fig.~\ref{fig2}(c).  Figures~\ref{fig2}(b) and~\ref{fig2}(c) have some resemblance, which one may attribute to similar differences in the quantum defects of the respective involved states. In the $n_i D \leftrightarrow n_f F$ series, the $\delta_{L, V}$ are much larger and the $M_V$ much smaller for $\Delta n \ge 0$ than for $\Delta n < 0$. Investigating further, we find that the $\Delta n \ge 0$ cases all show a Cooper-minimum-like behavior, similar to the $nD \leftrightarrow nP$ series in Fig.~\ref{fig2}(b). In addition, the EDTMs of the $n_i D \leftrightarrow n_f F$ series with $\Delta n > 0$ pass through maxima at series-dependent $n_i$-values located below the respective ``Cooper-minima''. These effects result in irregular appearances of the distribution of data points on the $n_i D \leftrightarrow n_f F$ series with $\Delta n \ge 0$ in Fig.~\ref{fig2}(c). 

For $\ell \ge 3$, the utilized model potentials become $\ell$-independent, and numerical readings of 
$\delta_{L,V} \ne 0$ are only due to numerical error. 
For instance, for $n_i F \leftrightarrow n_f G$ transitions (not shown here) we find $\delta_{L,V}$-values peaking near $10^{-6}$. This indicates a relative numerical confidence level of the EDTMs below $\sim 10^{-5}$. This was verified by comparing several results from Fig.~\ref{fig2} with similar results obtained with a finer spatial grid in the wave-function calculations.
This estimation of the numerical error is important because it shows that the  deviations between $M_L$ and $M_V$, discussed in Figs.~\ref{fig2}(a)-(c) and in the corresponding text passages above, are indeed due to the different gauge forms used. 

\begin{figure}[t!]
 \centering
  \includegraphics[width=0.48\textwidth]{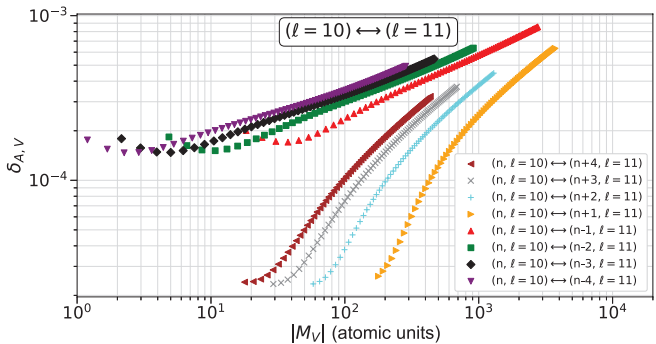}
  \caption{(Color online) Comparison of EDTMs calculated in AG vs VG using Eqs.~\ref{MV} and~\ref{MA}, respectively, for the indicated transition series with $\ell_{i/f} = 10$ or 11, for which gauge errors in the AG calculation are still significant. The  principal quantum number $n$ ranges between 15 (left) and 100 (right margins).}
  \label{fig3}
\end{figure}

Finally, we show a comparison between EDTMs computed in AG and VG. This comparison is of some fundamental interest but not relevant in applications of EDTMs. For low-$\ell$ states, $\delta_{A,V}$ can reach very large values (up to $\sim 10^6$, not shown here), reflecting the basic unsuitability of the AG for calculating EDTMs in non-Coulombic potentials. However, if $\ell$ becomes sufficiently large, the AG results tend to gradually approach the EDTMs calculated in VG and LG. As an example, in Fig.~\ref{fig3} we show results for several series for transitions between $\ell = 10$ and $\ell = 11$ states. On the series shown, the AG's gauge error is small but it still exceeds the numerical error by several orders of magnitude. It is noteworthy that there is no apparent (simple) relationship between $\delta_{A,V}$ and $|M_V|$, and that 
$\delta_{A,V}$ increases with $|M_V|$. Both of these observations qualitatively differ from Fig.~\ref{fig2}. In additional analyses, not presented, it is seen that $\delta_{A,V}$ drops below our numerical error for $\ell \gtrsim 20$ (with a caveat discussed in Sec.~\ref{sec:disc}). This shows that the always-present error of the AG that is caused by the long-range ion-core polarization potential, $-\alpha_d/(2 r^4)$, drops below our present numerical error when $\ell \gtrsim 20$. 

\maketitle
\section{Discussion}
\label{sec:disc}

In Sec.~\ref{sec:results} we have established and quantified deviations between LG- and VG- computations of EDTMs for transitions that involve Rydberg levels with $\ell_< \le 2$. These deviations are caused by the dependence of the utilized model potentials~\cite{Marinescu} on $\ell$ for $\ell \le 3$. For transitions with $\ell_< \ge 3$, the model potentials are identical, and the EDTMs in LG and VG agree within our numerical error of $\delta_{L,V}$ of $< 10^{-5}$. The gauge error of the AG, caused by the long-range, non-Coulombic core polarization potential $-\alpha_d / (2r^4)$, rapidly drops with increasing $\ell_<$ and becomes undetectable for $\ell_< \gtrsim 20$. Due to the ubiquitous use of low-$\ell$ Rb Rydberg states in science and technology, the following discussion is focused on aspects of the gauge errors of the LG pertaining to low-$\ell$ states.

For the strongest transitions within the $\ell_< \le 2$-series displayed in Fig.~\ref{fig2}, the choice of the gauge form does not play an overwhelming role, as $\delta_{L,V}$ - the relative EDTM error incurred by adopting the LG instead of the VG - barely exceeds $10^{-3}$ even in the worst cases. Picking $n_i = 40$ as a specific example, we find $\delta_{L,V} = 4.0 \times  10^{-5}$ for the strongest $S \leftrightarrow P$ transitions, $\delta_{L,V} = 1.7 \times 10^{-4}$ for the $40D \leftrightarrow 41P$ transition, and $\delta_{L,V} = 3.6 \times 10^{-5}$ for the $40D \leftrightarrow 39F$ transition. In high-precision spectroscopy, it would be fairly challenging to probe Rabi-frequency discrepancies at that level of precision (and to even calibrate RF electric fields with sufficient accuracy to enable such measurements). However, the weak $n_i P \leftrightarrow n_f D$ series and several of the $n_i D \leftrightarrow n_f F$ series have $\delta_{L,V}$-values in the 1$\%$-and-above range and exhibit several unusual behaviors, including Cooper-minimum-like features. The properties of these weak series may offer opportunities to explore gauge effects in Rydberg electric-dipole transitions for fundamental investigations. The same holds for photo-ionization cross sections, which generally show larger differences between LG and VG~\cite{Viray2021, Cardman2021} than bound-bound EDTMs.   

It is abundantly evident in Fig.~\ref{fig2} that all data points for any given series that involves $\ell$-dependent model potentials fall on approximately straight lines, regardless of any added complexity such as Cooper-like minima in $|M_V|$.  Fitted  slopes are $\approx -1$ for all three types of transitions in Fig.~\ref{fig2}. This observation implies that the absolute difference in EDTMs, $\Delta_{L,V} = M_L - M_V$, remains constant over a wide range of $n$. In all series presented in Fig.~\ref{fig2}, the relative variation of  $\Delta_{L,V}$ across the entire series is less than 2$\%$. Further investigation might be needed to explain this finding. The value of $\Delta_{L,V}$
is largest for the strongest series, $nD \leftrightarrow (n+1)P$, at $\Delta_{L,V} = -0.34$~ea$_0$.

The $1/\omega^2$-dependence of $M_A$ (see Eq.~\ref{MA}) and the $1/\omega$-dependence of $M_V$ (see Eq.~\ref{MV}) cause an issue of numerical accuracy for near-degenerate transitions. The integral expressions in the respective equations can drop to near or even below the numerical error. As a result, for near-degenerate transitions the expressions in Eqs.~\ref{MV} and~\ref{MA} can become numerically unstable and lead to non-physical results. (This is why in Fig.~\ref{fig3} we have excluded the result for $\Delta n = 0$.) Since the LG (Eq.~\ref{ML}) has no equivalent numerical problem, the frequency-denominator issue is of no practical concern. For $\ell_< \le 2$, where the VG {\sl{must}} be used, there are no near-degenerate transitions, so the issue does not arise, while for $\ell_< \ge 3$, where near-degenerate transitions occur when $\Delta n = 0$, the LG may be used. 

Following the discussion, we are using the method summarized in Fig.~\ref{fig1} to pick EDMTs for any electric-dipole transition of interest. First, we compute databases $M_V (n_i, l_i, \Delta n, \Delta_\ell)$, $M_L (n_i, l_i, \Delta n, \Delta_\ell)$ and $M_A (n_i, l_i, \Delta n, \Delta \ell)$. (The database $M_A$ is for added insight only.) After entering the transition labels $(n_i, l_i, \Delta n, \Delta \ell)$, the following simple rules are applied: 
\begin{itemize}
    \item If $\ell_< \le 2$, pick $M_V (n_i, l_i, \Delta n, \Delta \ell)$.
    \item If $\ell_< \ge 3$ and $|\Delta n| \ge 1$, use the average of $M_V (n_i, l_i, \Delta n, \Delta \ell)$ and $M_L (n_i, l_i, \Delta n, \Delta \ell)$.
    \item If $\ell_< \ge 3$ and $|\Delta n| =0 $, pick $M_L (n_i, l_i, \Delta n=0, \Delta \ell)$.
\end{itemize}

\maketitle
\section{Conclusion}
\label{sec:concl}
We have presented results for EDTMs calculated in three different gauge forms: velocity, length and acceleration. Based on the analysis of the observed differences between these forms, we outlined a method for choosing the most appropriate gauge, depending on the quantum numbers of the transition. We have discussed aspects of the underlying physics. In the computations we have used model potentials for Rb from~\cite{Marinescu}. Analogous computations could be performed for cesium or other alkali atoms for which there is a set of model potentials.  

The observed differences between the forms have negligible or only minor consequences in applications of Rydberg atoms for microwave field sensing~\cite{Sedlacek2012, Holloway2014, RFMS, anderson2021}, where strong transitions are used. For these, the relative effect of gauge on the EDTMs, $\delta_{L,V}$, remains below $\sim 10^{-3}$. Gauge effects with a significance $>1\%$ may be detectable in weak Rydberg-Rydberg transitions, which one may study to experimentally explore gauge effects. 

It is finally noted that the DC Stark effect occurs at $\omega = 0$, while the present work deals with resonant transitions at frequencies that are typically in the radio-frequency and microwave ranges. A limiting case of $\omega \rightarrow 0$, covered in our work, occurs for near-degenerate $\Delta n =0$ transitions between high-$\ell$ states. The limit $\omega \rightarrow 0$ requires caution because of the rise of Bloch-Siegert shifts caused by the counter-rotating terms~\cite{berman}, which we have dropped early-on when making the rotating-wave approximation, and because of other complications. The limit $\omega \rightarrow 0$ may be treated better as a quasi-static DC Stark effect in a slowly-varying electric field. In that approach, no resonance condition is assumed and the static atom-field interaction operator $e {\bf{E}} \cdot \hat {\bf {r}}$ applies, for which the length-gauge EDTMs are exact. Hence, the (quasi-static) DC quadratic Stark effect of low-$\ell$ atomic states requires the length-gauge form of the EDTMs, ({\sl{even}} at low $\ell$), whereas interactions with microwave electric fields involving the same low-$\ell$ states require the velocity-gauge form. This difference may become relevant in precision measurements of resonant AC versus DC electric fields as well as applications of Rydberg atoms in defining voltage standards~\cite{HollowayV2021}.

\maketitle
\section*{Acknowledgments}
\label{sec:acknowledgments}
We thank Ryan Cardman, Bineet Kumar Dash, and Dr. David Anderson for useful discussions. This work was supported by the NSF Grant No. PHY-2110049.

\bibliography{references.bib}

\end{document}